\documentclass[final,5p,times,twocolumn]{elsarticle}

\usepackage{amssymb}
\usepackage{amsmath}

\begin{document}

\begin{frontmatter}



\title{Computational models of sound-quality metrics using method for calculating loudness with gammatone/gammachirp auditory filterbank}


\author[1]{Takuto Isoyama}
\ead{isoyama-t@jaist.ac.jp}

\author[1]{Shunsuke Kidani}
\ead{kidani@jaist.ac.jp}

\author[1]{Masashi Unoki}
\ead{unoki@jaist.ac.jp}
\affiliation[aaa]{
organization={School of Information Science, Japan Advanced Institute of Science and Technology},
addressline={1-1 Asahidai},
city={Nomi},
postcode={923-1292},
state={Ishikawa},
country={Japan}
}

\begin{abstract}
Sound-quality metrics (SQMs), such as sharpness, roughness, and fluctuation strength, are calculated using a standard method for calculating loudness (Zwicker method, ISO532B, 1975). Since ISO 532 had been revised to contain the Zwicker method (ISO 5321) and Moore-Glasberg method (ISO 532-2) in 2017, the classical computational SQM model should also be revised in accordance with these revisions. A roex auditory filterbank used with the Moore-Glasberg method is defined separately in the frequency domain not to have impulse responses. It is therefore difficult to construct a computational SQM model, e.g., the classical computational SQM model, on the basis of ISO 532-2. We propose a method for calculating loudness using the time-domain gammatone or gammachirp auditory filterbank instead of the roex auditory filterbank to solve this problem. We also propose three computational SQM models based on ISO 532-2 to use with the proposed loudness method. We evaluated the root-mean squared errors (RMSEs) of the calculated loudness with the proposed and Moore-Glasberg methods. We then evaluated the RMSEs of the calculated SQMs with the proposed method and human data of SQMs. We found that the proposed method can be considered as a time-domain method for calculating loudness on the basis of ISO 532-2 because the RMSEs are very small. We also found that the proposed computational SQM models can effectively account for the human data of SQMs compared with the classical computational SQM model in terms of RMSEs.
\end{abstract}

\begin{highlights}
\item A method for calculating loudness using the time-domain gammatone/gammachirp auditory filterbanks is proposed.
\item Three computational SQM models of sharpness, roughness, and fluctuation strength are also proposed using the proposed loudness method.
\item The results of an evaluation indicate improvement in the estimation accuracy of metrics such as sharpness, roughness, and fluctuation strength with the proposed method.
\end{highlights}

\begin{keyword}
Loudness \sep Sound-quality metrics \sep Sharpness \sep Roughness \sep Fluctuation strength \sep Gammatone auditory filterbank \sep Gammachirp auditory filterbank
\end{keyword}
\end{frontmatter}

\section{Introduction}
There has been considerable interest in sound-quality metrics (SQMs) \cite{Zwicker}, such as sharpness, roughness, and fluctuation strength, which objectively examine the texture of sound as perceived by humans. These metrics have been applied in a variety of studies, including on sensory pleasantness \cite{Aures}, annoyance \cite{Zwicker}, product sound design \cite{Nykanen,Kwon}, and soundscape analysis \cite{Lionello,Mian}. SQMs play an important role in creating more desirable products and ambient sound. These metrics can be obtained by computational modeling. Metrics based on time variabilities of loudness, such as roughness and fluctuation strength, are commonly used as the standard method for calculating loudness using a time-domain auditory filter that was proposed by Zwicker (ISO 532B:1975).

ISO 532 was revised in 2017 as ISO 532-1:2017 \cite{5321} and ISO 532-2:2017\cite{5322}.
The methods for calculating loudness described in the former and the latter standards are called the Zwicker method and Moore-Glasberg method, respectively. The Zwicker method calculates loudness for stationary and time-varying sounds. The Moore-Glasberg method calculates only the loudness for stationary sounds. There are differences between these methods for calculating loudness such as the frequency scale (Bark scale or equivalent rectangular bandwidth (ERB) scale \cite{Moore}) and the auditory filter shape (symmetric or asymmetric) depending on the signal level.

A computational SQM model needs to be modified to match the revision of ISO 532. It is believed that minor modifications to this model would enable the computation of SQMs using the Zwicker method. The study of the computational SQM model using the Moore-Glasberg method is limited to sharpness \cite{sharpnessERB}. A roex auditory filterbank used with the Moore-Glasberg method is defined separately in the frequency domain to not have impulse responses. It is therefore difficult to construct a computational SQM model, e.g., the classical computational SQM model, obtained from the time variability of loudness such as roughness and fluctuation strength based on ISO 532-2:2017. If it is possible to compute SQMs based on ISO 532-2:2017, the ERB scale and the asymmetry of the auditory filter shape can be used to better estimate SQMs.

As a preliminary study, we constructed a model for calculating loudness using the time-domain gammatone auditory filterbank (GTFB) \cite{Gammatone} or analytical gammachirp auditory filterbank (GCFB) \cite{Gammachirp}, to determine if we could calculate sharpness and fluctuation strength using that model \cite{GTLoudnessFA,GTLoudness}. As a result, the calculated sharpness and fluctuation strength showed similar trends to the human data.

In this paper, we propose a method for calculating loudness using the time-domain GTFB or GCFB instead of the roex auditory filterbank to solve this problem. We also propose three computational SQM models that are based on ISO 532-2 to use with the proposed method.

This paper is organized as follows. Section 2 reviews related literature, Section 3 describes the proposed method for calculating loudness, Section 4 describes the proposed computational SQM models, Section 5 discusses the evaluation of the proposed models through comparison with other related computational SQM models, and Section 6 concludes the paper.

\section{Literature review}
\subsection{Method for calculating loudness}
ISO 532A:1975 is Stevens’ method for calculating the loudness level for broadband sounds. This method calculates the loudness level for sounds analyzed in three octave bands (1 octave, 1/2 octave, and 1/3 octave) according to Stevens’ law. However, this calculation method cannot account for loudness at low sound-pressure levels.

ISO 532B:1975 is a method for calculating loudness using the time-domain 1/3-octave filterbank as an approximated critical band filterbank for stationary sounds proposed by Zwicker. It is obtained by (1) correcting for the sound field and transferring the characteristics of the outer and middle ear, (2) calculating the excitation pattern for each critical band, (3) calculating the specific loudness, and (4) summing the specific loudness. The excitation pattern for each critical band is obtained by bandwidth division using a 1/3-octave filterbank constructed in accordance with the Bark scale. This calculation method incorporates the idea of partial loudness to correctly account for loudness even at low sound-pressure levels.

The Zwicker method \cite{5321} is an improved version of ISO 532B:1975 that can also be calculated for time-varying sounds. The basic flow of the calculation is the same as in ISO 532:B. There are five updates: (i) refinement of the auditory filterbank using a 1/3-octave filter, (ii) improved calculation of specific loudness from the excitation to internal noise ratio, (iii) refinements of the relationship between loudness and loudness level, (iv) addition of nonlinear time-decay processing for auditory, and (v) second-order leakage integral process added to the sum of the specific loudness.

The Moore-Glasberg method \cite{5322} calculates loudness using the frequency-domain auditory filterbank for monaural and binaural sounds. This section describes this method for monaural sound. The basic flow of the calculation is the same as in ISO 532:B. There are differences between the Zwicker and Moore-Glasberg methods such as the frequency scale (Bark scale or ERB scale \cite{Moore}) and auditory filter shape (symmetric or asymmetric) depending on the signal level.

\subsection{Computational model of SQMs}
Bismarck established the correlation between specific loudness and sharpness and subsequently presented a computational model of sharpness using ISO 532B:1975 \cite{Bismarck, Zwicker}. Fastl and Zwicker generalized Bismarck’s model \cite{Zwicker}. Aures modified Bismarck’s proposed computational model of sharpness to a loudness-dependent model \cite{Aures}. Swift and Gee proposed a computational model of sharpness from the specific loudness calculated using the Moore-Glasberg method \cite{sharpnessERB}.

Aures proposed a computational model of roughness on the basis of the relationship between modulation depth and roughness for different critical bands \cite{AuresR}. This model was optimized by Daniel and Weber to match human perception \cite{DanielR}. A computational model of roughness based on the ERB scale was also proposed by Duisters \cite{DuistersR}. This model is more consistent with the human data of roughness than Daniel and Weber’s model. Widmann and Fastl found that the difference between the peaks and valleys of the temporal masking pattern is related to the perception of roughness \cite{FastlR} and proposed a computational model of roughness using the temporal variation of specific loudness, which is the output of the Zwicker method for calculating loudness \cite{Widmann}.

Vecchi proposed a computational model of fluctuation strength on the basis of the relationship between fluctuation strength and modulation frequency \cite{Vecchi}. Fastl clarified the relationship between the temporal masking pattern and fluctuation strength as well as roughness and proposed a computational model of fluctuation strength \cite{FastlF}.

Fastl’s study has made it possible to consistently consider SQMs using a method for calculating loudness \cite{Zwicker,FastlR,FastlF}. This indicates that loudness is important in the computation of SQM. The use of the ERB scale improves the SQM computation results, as suggested by Duisters \cite{DuistersR}.

\subsection{Research issue}
With the revision of ISO 532, a computational SQM model also needs to be updated. Since the Zwicker method calculates loudness using the time-domain auditory filter similar to ISO 532B:1975, no major updates are expected.

The study of computational SQM models using this Moore-Glasberg method is limited to sharpness, and it is difficult to construct a computational SQM model for roughness and fluctuation strength obtained from the time variation of loudness. A frame-by-frame loudness-calculation method is also possible, but it requires the use of an auditory filter with an impulse response for loudness calculation because it is difficult to capture minute changes in loudness.

If we can construct a computational SQM model on the basis of ISO 532-2:2017, the following two points are possible. (a) Compared with the filterbank based on the Bark scale the filterbank based on the ERB scale has a higher resolution in the low-frequency band. Therefore, the accuracy of SQM estimation in the low-frequency range may be improved in the calculation of SQMs. (b) Since the asymmetry of the auditory filter shape depends on the sound-pressure level, the accuracy of SQM estimation with respect to changes in sound-pressure level may be improved.
 \begin{figure*}[!tb]
  \center
  \includegraphics[width=1\textwidth]{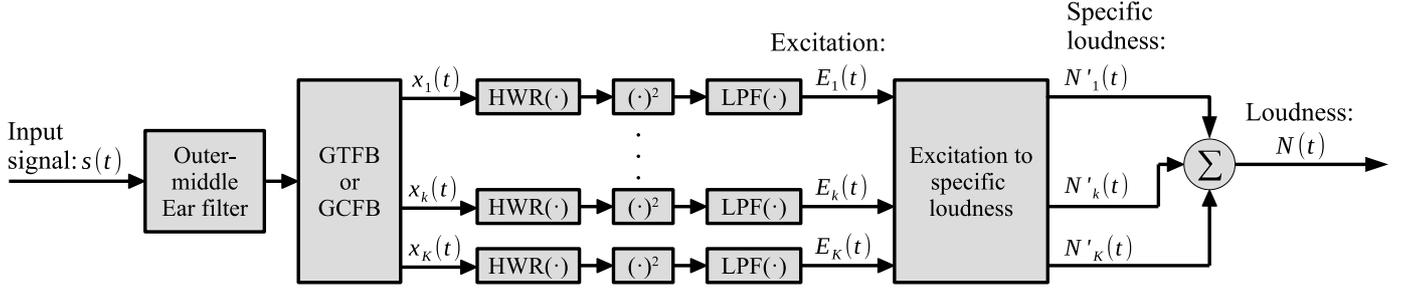}
  \caption{Block diagram of proposed method for calculating loudness (proposed loudness method) using gammatone auditory filterbank (GTFB) or gammachirp auditory filterbank (GCFB).}
  \label{fig:LoudnessModel}
\end{figure*}
\section{Proposed method for calculating loudness}
\label{Chap:3}
Instead of the roex auditory filterbank, we used the GTFB or GCFB as the time-domain auditory filterbank to develop our method for calculating loudness, and the method was used to construct our computational SQM models.

Figure \ref{fig:LoudnessModel} shows a block diagram of the proposed method. The proposed method using the GTFB is called the GT loudness method, and it using the GCFB is called the GC loudness method. The input signal $s(t)$ is first filtered to mimic the characteristics of the outer and middle ear then divided into $K$ frequency channels $x_k(t)$ using the GTFB or GCFB. Next, excitation $E_k(t)$ is calculated from the divided signal $x_k(t)$. Finally, loudness $N(t)$ is obtained by summing the specific loudness $N'_k(t)$ calculated from $E_k(t)$. The HWR , $(\cdot)^2$, and ${\rm{LPF(\cdot)}}$ in the figure denote half-wave rectification, squaring, and low-pass filtering, respectively.

\subsection{Auditory filterbank}
\subsubsection{Gammatone filterbank}
The GTFB is constructed using the impulse response of the gammatone auditory filter function \cite{Gammatone} defined as
\begin{equation}
  gt_k(t) = at^{(M-1)}\exp{(-2\pi b{\rm{ERB_N}}(f_k)t)}\cos{(2 \pi f_kt+\phi)},
  \label{Eq:gt_imp}
\end{equation}
where $a$, $t$, $M=4$, $b=1.019$, $f_k$, and $\phi$ are amplitude, time, the order of gammatone function, the constant, center frequency of the $k$-th filter, and phase, respectively.

The equivalent rectangular bandwidth (${\rm{ERB_N}}$) and $f_k$ are defined as
\begin{align}
  {\rm{ERB_N}}(f_k) =& 24.7(4.37f_k/1,000+1),\\
  f_k =& \left(10^{\frac{{\rm{ERB_{N}\mathchar`-number}}}{21.4}}-1\right)\frac{1,000}{4.37},
\end{align}
where the subscript $N$ indicates that it was derived from an experiment on a normal-hearing person.

The GTFB uses the $4$-th cascade one-zero two-pole gammatone proposed by Slaney \cite{IIRGammatone}. The GTFB is arranged such that the ${\rm{ERB_{N}\mathchar`-number}}$ lines up from 1.8 Cam ($k=1$) to 38.9 Cams ($k = K = 372$) in 0.1-Cam increments, similar with the Moore-Glasberg method \cite{5322}.

\subsubsection{Gammachirp filterbank}
The analytical gammachirp filter accounts for the asymmetry of the auditory filter shape in the GTFB and proposed by Irino and Patterson \cite{Gamma}. The compressive gammachirp auditory filter \cite{roexgcfb,CGammachirp} and dynamic-compressive gammachirp auditory filter \cite{DCGammachirp} were also been proposed for this auditory filterbank to account for compression properties. The Moore-Glasberg method accounts for the asymmetry of the auditory filter shape but does not for the compression property. To carry out the same process as ISO 543-2, the proposed method uses the analytical gammachirp filter.

The GCFB is constructed using the impulse response of the analytical  gammachirp auditory filter with real coefficients defined as
\begin{equation}
      gc_k(t) = at^{(M-1)}\exp{(-2\pi b{\rm{ERB_N}}(f_k)t)}\cos{(2 \pi f_kt+c\ln{t}+\phi)}, \label{Eq:gc_imp}
\end{equation}
where $c$ denotes the coefficient of frequency change (chirp) \cite{Gamma} and $\ln$ denotes the natural logarithmic operator. The only difference between the gammachirp auditory filter and impulse response of the gammatone auditory filter (Eq. (\ref{Eq:gt_imp})) is the chirp term $c\log{t}$. When $c=0$, the chirp term ($c\ln{t}$) vanishes and has the same frequency characteristics as the gammatone auditory filter. By transforming Eq. (\ref{Eq:gc_imp}) into a complex impulse response and applying Fourier transform, the frequency characteristics of the gammachirp auditory filter can be represented as
\begin{align}
|G_c(f)| =& \frac{|\Gamma(M-jc)|\exp{(c\theta(f))}}{|2\pi\sqrt{(b{\rm{ERB_N}}(f_k))^2+(f-f_k)^2}|},\\
 =&  a_\Gamma|G_T(f)|\exp{(c\theta(f))},\\
\theta(f)  =&  \arctan\left(\frac{f-f_k}{b{\rm{ERB_N}}(f_k)}\right),
\end{align}
where $f$ denotes frequency, $a_\Gamma$ denotes amplitude, and the amplitude spectrum of the gammatone auditory filter $|G_T(f)|$ is asymmetric on a linear frequency axis. Since $\theta(f)$ has asymmetry around $f_k$, $\exp{(c\theta(f))}$ is an asymmetric function. When $c$ is negative, $\exp{(c\theta(f))}$ becomes a low-pass filter (LPF), and when $c$ is positive, $\exp{(c\theta(f))}$ becomes a high-pass filter. This enables control of asymmetry in the low/high range of the gammachirp auditory filter. By making $c$ a function of sound-pressure level \cite{Gammachirp}, as shown in the following equation, level dependence and asymmetry of the auditory filter can be added.
\begin{equation}
  c = 3.38 - 0.107Ps_k,
\end{equation}
where $Ps_k$ denotes the sound-pressure level resulting from the output of the different filters in the GTFB. The $c$ is smoothed in the ${\rm{ERB_{N}\mathchar`-number}}$ direction using a weighted moving average.

The $\exp{(c\theta(f))}$ of the current gammachirp filter is designed by cascading the minimum phase filter of the infinite-impulse-response (IIR) filter \cite{Gammachirp}. The GCFB is arranged such that the ${\rm{ERB_{N}\mathchar`-number}}$ lines up from 2.6 Cam ($k=1$) to 36.9 Cam ($k=K=344$) in 0.1-Cam increments.

\subsection{Calculation of excitation}
To simulate the response of the inner hair cells and auditory nerve, $E_k(t)$ is calculated by HWR, ($\cdot$)2, and leakage integration processing through an LPF of the output of GTFB or GCFB. The transfer characteristic of the leakage integrator $H_{{\rm{LPF}}}(\omega)$ is defined as a second-order LPF in which two LPFs shown in the following equations are cascaded.
\begin{align}
  H_{{\rm{LPF}}}(\omega) &= \frac{a_{\rm LPF}}{1-\exp{(-2\pi f_c/f_s)}\exp{(j\omega)}},\\
  a_{\rm LPF}&=\frac{1}{1-\exp{(-2\pi f_c/f_s)}},
\end{align}
where $fs$ denotes sampling frequency and is $44,100$ Hz, $\omega$ denotes the angular frequency ($2\pi f$), and $fc$ is cut-off frequency of the leakage integrator and is $1,200$ Hz. An $a_{\rm LPF}$ is the amplitude of the leakage integrator.

\subsection{Calculation of loudness from excitation}
The $N'_k(t)$ is calculated from $E_k(t)$ in accordance with three conditions:
\begin{itemize}
  \item Condition of $E_k(t)/E_0 < E_{{\rm THRQ},k}$
\begin{equation}
\begin{split}
      N_k'(t) =& Q_N\left(\frac{2E_k(t)/E_0}{(E_k(t)/E_0+E_{{\rm THRQ},k})}\right)^{1.5} \\
      &\times\biggl((GE_k(t)/E_0+A)^{\alpha}-A^{\alpha}\biggr),
\end{split}
\end{equation}
  \item Condition of $E_{{\rm THRQ},k} \leq E_k(t)/E_0 < 10^{10}$
  \begin{equation}
    N_k'(t) = Q_N\biggl((GE_k(t)/E_0+A)^{\alpha}-A^{\alpha}\biggr),
  \end{equation}
  \item Condition of $E_k(t)/E_0 > 10^{10}$
  \begin{equation}
    N_k'(t) = Q_N\left(\frac{E_k(t)/E_0}{0.99\cdot10^{-3}}\right)^{0.2}, \label{aaaa}
  \end{equation}
\end{itemize}
where $Q_N$ denotes the loudness coefficient and is $54.6\times10^{-3}$ when the GTFB is used and $54.8\times10^{-3}$ when the GCFB is used, $E_{{\rm THRQ},k}$ denotes the excitation corresponding to the minimum audible value in silence, $G$ denotes the low-level gain of cochlear amplification, $\alpha$ denotes the power exponent of Stevens’s power law, and $A$ denotes the input/output characteristic \cite{5322}. Since the shapes of the roex auditory filterbank specified with the Moore-Glasberg method are different from those of the GTFB or GCFB, the areas of the specific loudness calculated with these filterbanks are also different. Therefore, we added 0.049 for the GT loudness method and 0.047 for the GC loudness method to $\alpha$ ($=0.2$) to align the range of specific loudness of the Moore-Glasberg method and that of the proposed method. The $N(t)$ is obtained from the total computation of specific loudness defined as
\begin{equation}
  N(t) = \sum^K_{k=1}N_k'(t).
\end{equation}

\begin{figure}[!tb]
  \center
  \includegraphics[width=0.5\textwidth]{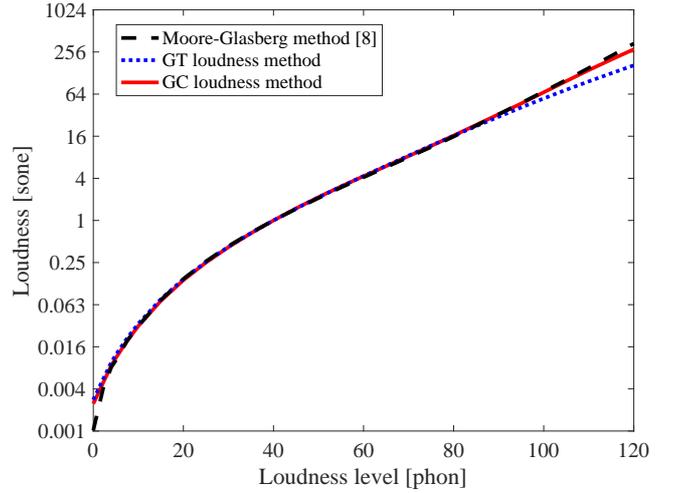}
  \caption{Relationship between loudness level and loudness}
  \label{fig:Loudlevel}
\end{figure}

Figure \ref{fig:Loudlevel} shows the relationship between loudness and loudness levels, where the horizontal axis represents the loudness level in phon and the vertical axis represents the loudness in sone. When the loudness level is greater than 80 phon, the loudness of the GT loudness method is calculated to be lower than that of the Moore-Glasberg method. However, the loudness of the GT loudness method is consistent with the Moore-Glasberg method at 80 phon or less. The loudness of the GC loudness method are almost the same as those of the Moore-Glasberg method.

\begin{figure*}[!tb]
  \center
  \includegraphics[width=0.7\textwidth]{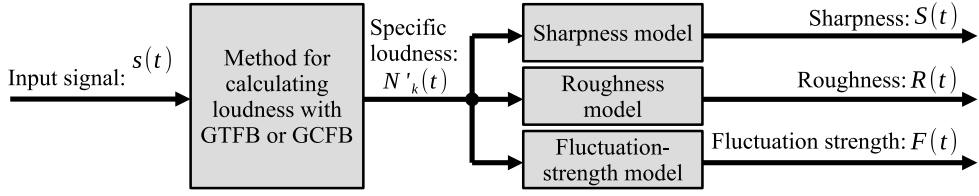}
  \caption{Block diagram of proposed computational sound-quality metric (SQM) models}
  \label{fig:SQMModel}
\end{figure*}

\section{Proposed computational model of SQMs}
Figure \ref{fig:SQMModel} shows the block diagram of the proposed SQM models \footnote{Software Public URL:\\ https://jstorage-2018.jaist.ac.jp/s/8Wk6tgt4LS6YTDo\\Password: Qx4AjwZwEi}. The $N(t)$ and $N'(t)$ are calculated from the observed $s(t)$ by using the proposed method. Sharpness is then computed by calculating the weighted center of the obtained $N'(t)$. Roughness and fluctuation strength are determined by analyzing $N'(t)$.
\subsection{Proposed sharpness model}
The proposed computational SQM model for sharpness (hereafter, proposed sharpness model) is designed on the basis of Aures’s computational model of sharpness \cite{Aures}, which takes loudness dependence into account.

\begin{figure}[!tb]
  \center
  \includegraphics[width=0.5\textwidth]{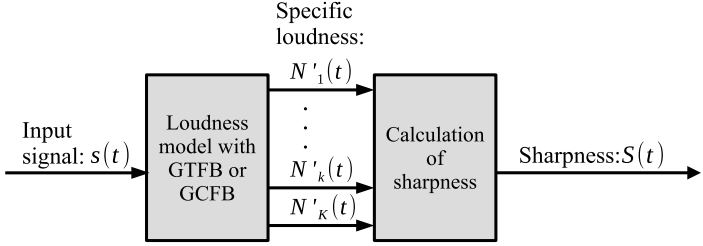}
  \caption{Block diagram of proposed sharpness model}
  \label{fig:sharpness}
\end{figure}
Figure \ref{fig:sharpness} shows a block diagram of the proposed sharpness model using the GTFB or GCFB. This model using the GTFB is called the GT sharpness model, and it using the GCFB is called the GC sharpness model. Sharpness $S(t)$ is obtained by calculating the weighted center from $N'_k(t)$ defined as
\begin{align}
S(t) =& Q_s\frac{\sum^K_{k=1}q_{{\rm s},k}(t)N'_k(t){\rm{ERB_N\mathchar`-number}}}{\sum^K_{k=1}N'_k(t)},\\
q_{{\rm s},k}(t) =& \frac{w_{{\rm s},k}\sum^K_{k=1}N'_k(t)}{{\rm{ERB_{N}\mathchar`-number}}\times\ln\left(\frac{N(t)+20}{20}\right)},
\end{align}
where $Q_s$ denotes the sharpness coefficient and is $2.29\times10^{-3}$ when the GTFB is used and $2.23\times10^{-3}$ when the GCFB is used, $q_{{\rm s},k}(t)$ denotes a weight that varies depending on loudness, and $w_{{\rm s},k}$ denotes the weighting function. The $w_{{\rm s},k}$ is fitted to minimize the root-mean-square error (RMSE) between the human data of sharpness \cite{Zwicker} and the output of this model. The $w_{{\rm s},k}$ is defined as
\begin{equation}
\begin{split}
w_{{\rm s},k}=&1.19\times10^{-3} {\rm{ERB_{N}\mathchar`-number}}^3\\
&-4.90\times10^{-2}{\rm{ERB_{N}\mathchar`-number}}^2\\
&+7.17\times10^{-1}{\rm{ERB_{N}\mathchar`-number}}\\
&-2.01.
\end{split}
\end{equation}

\subsection{Proposed roughness model}
Due to differences in the loudness model, Fastl’s model for computing roughness cannot be used directly. This model was developed on the basis of Duisters’ roughness model \cite{DuistersR} but with modifications in the computations.

\begin{figure*}[!tb]
  \center
  \includegraphics[width=0.9\textwidth]{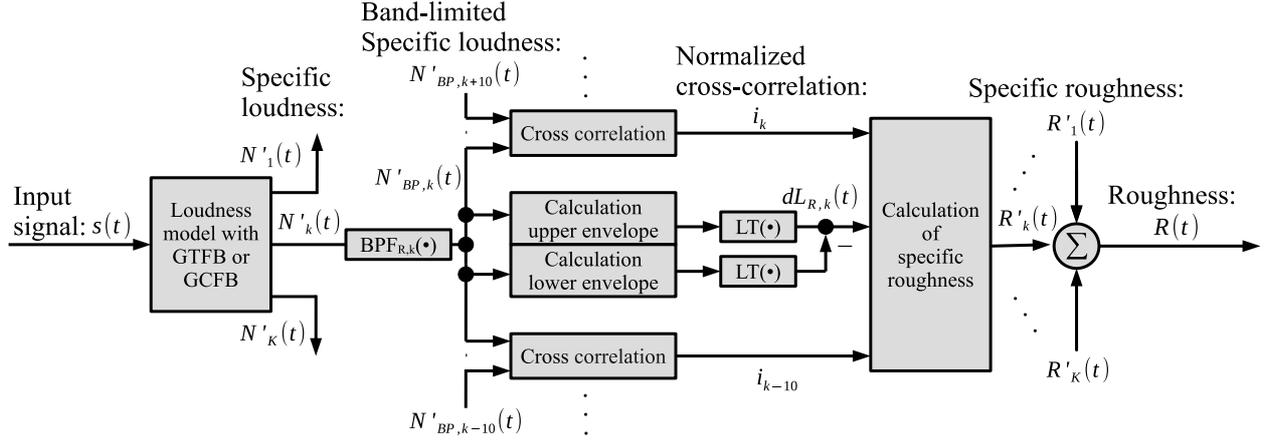}
  \caption{Block diagram of proposed roughness model}
  \label{fig:RoughnessModel}
\end{figure*}
\begin{figure}[!tb]
  \center
  \includegraphics[width=0.48\textwidth]{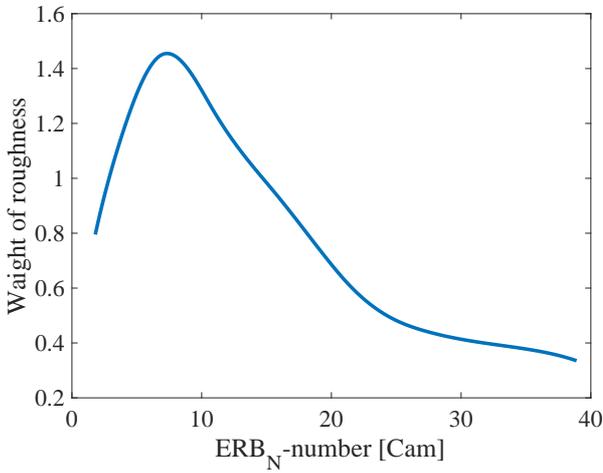}
  \caption{Weighting function of proposed roughness model}
  \label{fig:WFR}
\end{figure}

Figure \ref{fig:RoughnessModel} shows a block diagram of the proposed computational SQM model for roughness (hereafter, proposed roughness model). The proposed roughness model using the GTFB is called the GT roughness model, and it using the GCFB is called the GT roughness model. The proposed roughness model starts by obtaining $N'k(t)$ from the observed $s(t)$ using the proposed method. The bandwidth is then limited using a band-pass filter, and the upper and lower envelope of the band-limited specific loudness is obtained using a Hilbert transform and LPF. A logarithmic transformation is applied to these envelopes to obtain the difference between the peak and dip $\Delta L_{{\rm R},k}(t)$. The normalized cross-correlations $i_{k}$ and $i_{k-10}$ are calculated from the $k$-th band-limited loudness level and the $k\pm10$-th ($\pm1$ Cam) band-limited loudness level, respectively. The specific roughness $R'_k(t)$ is then calculated using $\Delta L_{{\rm R},k}$ and $i_k$ and $i_{k-10}$, and roughness $R(t)$ is obtained by computing its area. The following equations are used for the proposed roughness model.
\begin{equation}
  R = Q_{\rm R}\sum^K_{k=1}R'_k,
\end{equation}
\begin{equation}
R'_k =\\
\left\{
\begin{array}{ll}
\left(w_{{\rm R},k}\Delta L_{{\rm R},k}(t)i_k\right)^2
\hspace{3.8em} k \in [1,10], \\
\left(w_{{\rm R},k}\Delta L_{{\rm R},k}(t)(i_{k-10}i_k)\right)^2
\hspace{1em}k \in [11, K-11], \\
\left(w_{{\rm R},k}\Delta L_{{\rm R},k}(t)i_{k-10}\right)^2
\hspace{2.6em}k \in [K-10, K],
\end{array}
\right.
\end{equation}
where $Q_R$ denotes the roughness coefficient and is $3.15\times10^{-3}$ when the GTFB is used and $3.20\times10^{-3}$ when the GCFB is used and $(w_{{\rm R},k}$ denotes the weighting function optimized to match the human data of roughness \cite{Zwicker}, as shown in Fig. \ref{fig:WFR}.
\subsubsection{Difference between peak and dip in specific loudness}
\label{Cha:421}
The roughness model proposed by Duisters uses modulation depth as a parameter. Fastl discovered that the difference between the peaks and dips $\Delta L_{{\rm R},k}(t)$ of the log-transformed temporal masking pattern is related to roughness \cite{FastlR}. This enabled him to calculate roughness. The proposed roughness model incorporates this insight, and instead of using the modulation level, it log-transforms the specific loudness and uses $\Delta L_{{\rm R},k}(t)$, which is defined as
\begin{align}
\label{Eq:022}
  \Delta L_{{\rm R},k}(t) =& \Bigl(L'_{{\rm{R,Upper}},k}(t)-L'_{{\rm{R,Lower}},k}(t)\Bigl)W_{{\rm Calib},k}(t),\\
    W_{{\rm Calib},k}(t) =& \frac{{\rm LT}\Bigl(N'_k(t)\Bigl)}{\max\Bigl({\rm LT}(N'_k(t))\Bigl)},
\end{align}
where ${\rm LT}$ denotes a logarithmic transformation function created from the relationship between the loudness and loudness levels in Fig. \ref{fig:Loudlevel}. Missing data points are interpolated using a linear interpolation method to complete the function. The notation $W_{{\rm Calib},k}(t)$ is the weighting function that ensures that the maximum loudness level at each time is 1, and $N'_{{\rm{Upper}},k}(t)$ and $N'_{{\rm{Lower}},k}(t)$ represent the upper and lower envelopes of the band-limited specific loudness $N'_{{\rm{BP}},k}(t)$, respectively, as follows.
\begin{equation}
\begin{split}
  L'_{{\rm{R,Upper}},k}(t) =& {\rm LT}\Bigl({\rm LPF}(|N'_{{\rm{BP}},k}(t)+j{\rm Hilbert}(N'_{{\rm{BP}},k}(t))|)\Bigl),
\end{split}
\end{equation}
\begin{equation}
\begin{split}
  L'_{{\rm{R,Lower}},k}(t) =& -{\rm LT}\Bigl({\rm LPF}(|-N'_{{\rm{BP}},k}(t)+j{\rm Hilbert}(-N'_{{\rm{BP}},k}(t))|)\Bigl),
\end{split}
\end{equation}
where $H_{0,k}$ denotes the direct-current (DC) component of $N'_k(t)$. The LPF is a ninth-order IIR Butterworth LPF, with a cut-off frequency of 7 Hz.
\begin{figure*}[!tb]
  \center
  \includegraphics[width=0.9\textwidth]{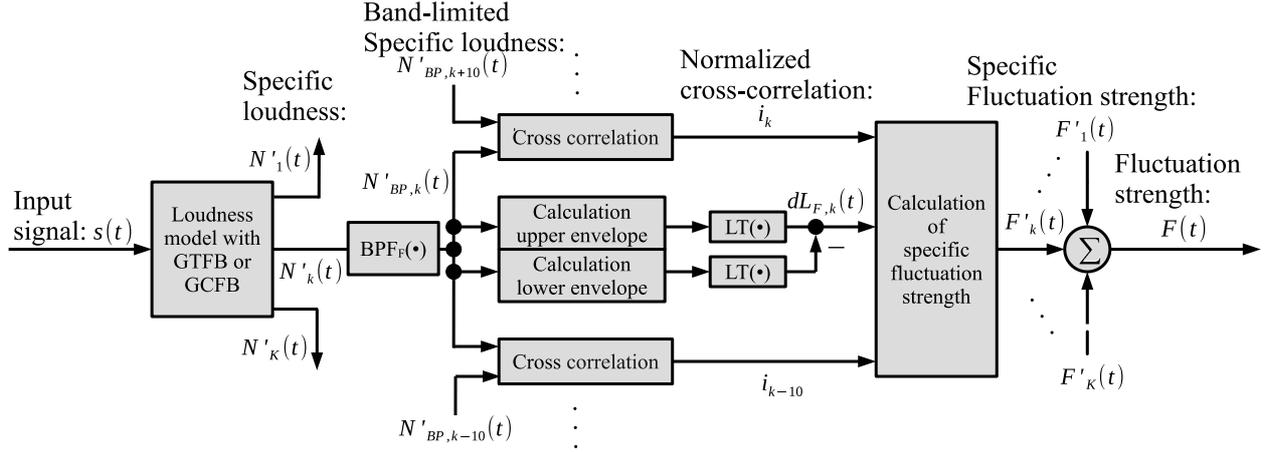}
  \caption{Block diagram of proposed fluctuation-strength model}
  \label{fig:fluctuation}
\end{figure*}
\subsubsection{Band-pass filter for proposed roughness model}
The $N'_{{\rm{BP}},k}(t)$ is defined as
\begin{equation}
N'_{{\rm{BP}},k}(t) ={\rm{BPF_{R,{\it k}}}}\Bigl(N'_k(t)-H_{0,k}\Bigl),
\end{equation}
where ${\rm{BPF_{R,{\it k}}}}$ is the band-pass filter optimized using a sigmoid function in accordance with the human data of roughness \cite{Zwicker}. The center frequency $C_{{\rm F},k}$ and band width $W_{{\rm B},k}$ of the band-pass filter are defined as
\begin{align}
C_{{\rm F},k}=&\frac{69.2}{1+\exp{\Bigl(-({\rm{ERB_N\mathchar`-number}}-\alpha)/\beta}\Bigl)},\\
W_{{\rm B},k}=&1.58C_{{\rm F},k},
\end{align}
where $\alpha=4.58$ and $\beta=1.48$ and constant. The band-pass filter used the gammatone filter defined as
\begin{equation}
    {\rm{BPF_{R,{\it k}}}}(t) = at^{(M-1)}\exp{(-2\pi W_{{\rm B},k}(C_{{\rm F},k})t)}\cos{(2 \pi C_{{\rm F},k}+\phi)}, \label{Eq:gt_imp2}
\end{equation}
where the filter order $M$ is 3.

\subsubsection{Calculation of normalized cross-correlation}
Even though the sense of variability is barely perceptible in sounds such as pink noise and white noise, roughness is high when these types of noise are the input. To sense this small variability, we calculate the normalized cross-correlation $i$ between distant auditory filters and incorporate it into the proposed roughness model. The $i$ is calculated from the time variability of $N'_{{\rm{BP}},k}(t)$ calculated between distant auditory filters defined as
\begin{equation}
i = \max\Bigl(\frac{\int_{\tau=-0.01}^{0.01}{x(t)y(t+\tau)}}{\sqrt{V_xV_y}}\Bigl),
\label{eq:cross}
\end{equation}
where $\tau$, $T$, and $V$ denote lag, signal length, and variance, respectively. The time variation of $N'_{{\rm{BP}},k}(t)$ calculated from the GTFB or GCFB has a time delay among the filters. Thus, we shift $i$ in 1-sample increments up to $10$ ms and set the maximum $i$ in channel $k$. Here, $i_k-10$ is the result of $i$ when $x$ is $N'_{{\rm{BP}},k-10}(t)$, and $y$ is $N'_{{\rm{BP}},k}(t)$, while $i_k$ is the result of $i$ when $x$ is $N'_{{\rm{BP}},k}(t)$, and $y$ is $N'_{{\rm{BP}},k+10}(t)$.

\subsection{Proposed fluctuation-strength model}
The process of calculating fluctuation strength is almost the same as that of computing roughness. The difference is the modulation frequency at which the sensation of fluctuation is felt. The proposed computational SQM model for fluctuation strength (hereafter, proposed fluctuation-strength model) was constructed using the proposed roughness model.

Figure \ref{fig:fluctuation} shows a block diagram of the proposed fluctuation-strength model. The model using the GTFB is called the GT fluctuation-strength model, and it using the GCFB is called the GC fluctuation-strength model. This model was developed with the following modifications.
\begin{align}
  F =& Q_{\rm F}\sum^K_{k=1}F'_k,
\end{align}
\begin{align}
F'_k =&
\left\{
\begin{array}{ll}
\Delta L_{{\rm F},k}^{0.6}(t)i_k^2
\hspace{4.3em}k \in [1, 10], \\
\Delta L_{{\rm F},k}^{0.6}(t)(i_{k-10}i_k)^2
\hspace{1em}k \in [11, K-11], \\
\Delta L_{{\rm F},k}^{0.6}(t)i_{k-10}^2
\hspace{2.9em}k \in [K-10, K],
\end{array}
\right.
\end{align}
where $Q_F$ denotes the fluctuation-strength coefficient and is $30.2\times10^{-3}$ when the GTFB is used and $30.0\times10^{-3}$ when GCFB is used. The $i_k$ and $i_{k-10}$ are obtained from Eq. (\ref{eq:cross}), as in the proposed roughness model.

\subsubsection{Difference between peak and dip in specific loudness}
The difference between the peaks and dips $\Delta L_{{\rm F},k}(t)$ is obtained as
\begin{align}
\label{Eq:024}
  \Delta L_{{\rm F},k}(t) =& \Bigl(L'_{{\rm{F,Upper}},k}(t)-L'_{{\rm{F,Lower}},k}(t)\Bigl)W_{{\rm Calib},k}(t),\\
    W_{{\rm Calib},k}(t) =& \frac{{\rm LT}\Bigl(N'_k(t)\Bigl)}{\max\Bigl({\rm LT}(N'_k(t))\Bigl)},
\end{align}
where ${\rm LT}$ is the same function described in Chapter \ref{Cha:421}. $N'_{{\rm{Upper}},k}(t)$ and $N'_{{\rm{Lower}},k}(t)$ are the upper envelope of band-limited specific loudness and the lower envelope of specific loudness as follows.
\begin{align}
  L'_{{\rm{F,Upper}},k}(t) =& {\rm LT}\Bigl({\rm LPF}(|N'_{{\rm{BP}},k}(t)+j{\rm Hilbert}(N'_{{\rm{BP}},k}(t))|)\Bigl),\\
  L'_{{\rm{F,Lower}},k}(t) =& -{\rm LT}\Bigl({\rm LPF}(|-N'_{{\rm{BP}},k}(t)+j{\rm Hilbert}(-N'_{{\rm{BP}},k}(t))|)\Bigl),
\end{align}
where the LPF is a ninth-order IIR Butterworth low-pass filter, with a cut-off frequency of 0.4 Hz.

\subsubsection{Band-pass filter for proposed fluctuation-strength model}
The $N'_{{\rm{BP}},k}(t)$ is defined as
\begin{equation}
N'_{{\rm{BP}},k}(t)={\rm{BPF_F}}\Bigl(N'_k(t)-H_{0,k}\Bigl)
\end{equation}
The band-pass filter consists of a second-order IIR Butterworth LPF and a second-order IIR Butterworth high-pass filter connected in cascade. The cut-off frequencies are 5 and 2 Hz, respectively.

\section{Evaluation}
We evaluated the RMSE of loudness calculation using the proposed method and the Moore-Glasberg method to see if the proposed method can calculate loudness correctly. We then evaluated whether the output of the proposed computational SQM models can account for the human data of the SQMs \cite{Zwicker} by calculating the RMSE between the model output and  results obtained from the human data of the SQMs. To determine the effectiveness of the proposed SQM models, we conducted a similar evaluation of conventional SQM models and compared these results with those of the proposed SQM models. Since the inputs of the proposed SQM models are assumed stationary signals, the average of the outputs of the proposed SQM models was used in the evaluation.
\begin{table*}[t]
  \center
  \caption{Sound signal used for evaluating proposed loudness method}
  \vspace{1em}
  \label{tab:001}
  \begin{tabular}{rcc}\hline
 Frequency [Hz]	& Sound pressure level [dB]	     & Step size [dB] \\ \hline\hline
 100         & 50		     & ---\\
 1,000       & 20 $\sim$ 80	 & 10\\
 3,000  		& 20 $\sim$ 80	 & 20\\\hline
\end{tabular}
\end{table*}
\subsection{Evaluation of loudness}
We compared the loudness calculated with the proposed method with the loudness specified in ISO 532-2. We evaluated the loudness for different frequency. Table \ref{tab:001} shows the input signal (sinusoidal signal) used for the loudness evaluation. The loudness with the proposed method was calculated using time-averaged loudness.

\begin{figure}[!tb]
    \begin{center}
        \includegraphics[width=0.5\textwidth]{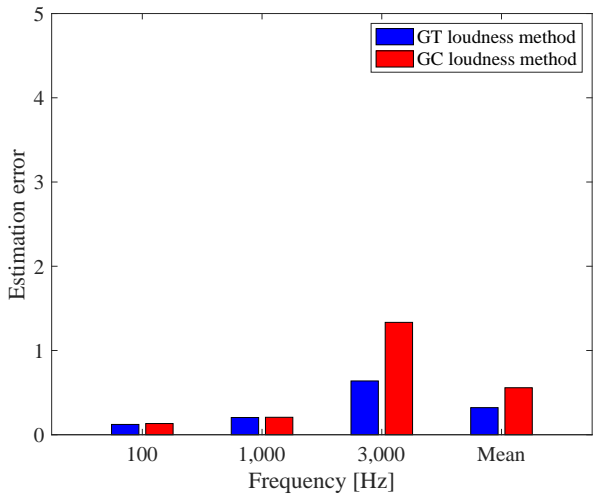}
        \caption{Estimation error of both versions of proposed loudness method: GT loudness method and GC loudness method}
        \label{fig:LoudnessResults}
    \end{center}
\end{figure}

Figure \ref{fig:LoudnessResults} show the results of the calculating loudness for different frequencies. The horizontal axis is the frequency and the vertical axis is the estimated error of the calculated results of the proposed loudness method relative to the calculated results of the Moore-Glasberg method. The results of the proposed method were in almost perfect agreement with the results of the Moore-Glasberg method. The estimation error at a frequency of 3,000 Hz was the largest, but since the loudness at this frequency was 27 sones, the estimation error is considered sufficiently small. This indicates that the calculation results of the proposed loudness method are consistent with those of the Moore-Glasberg method.

\begin{table*}[t]
    \caption{Sound signal used for evaluating proposed sharpness model}
    \label{tab:typenoise}
    \center
    \begin{tabular}{ccrcc}\hline
        Type of noise   & Center frequency [Hz]	& Bandwidth [Hz]     & Low frequency [Hz] &  High frequency [Hz] \\ \hline\hline
        NB noise        & 200 $\sim$ 10,000     & 104 $\sim$ 2,463      & ---		          & ---			         \\
        HP noise        & ---                   & 1,500 $\sim$ 9,750 & 250 $\sim$ 8,500	  & 10,000 (constant)	 \\
        LP noise    	& ---  			        & 150 $\sim$ 10,300   & 200 (constant)     & 350 $\sim$ 10,500	 \\\hline
    \end{tabular}
\end{table*}

\subsection{Evaluation of sharpness}
Both versions of the proposed sharpness model (GT and GC sharpness models) was compared with previous sharpness models, i.e., Fastl \& Zwicker’s model (FZ model) and the loudness-dependent Aures model (Aures model). Sharpness was evaluated for three different noise \cite{Zwicker} and loudness levels \cite{Aures}.

The initial set of stimuli comprised narrow-band (NB) noise, high-pass (HP) noise, and low-pass (LP) noise, as listed in Table \ref{tab:typenoise} and specified in DIN 45692 \cite{din45692}. The amplitudes of these sound stimuli were adjusted so that the loudness from the proposed loudness method was 4 sones by time-averaged sharpness. The second set of stimuli was a 0.5-sec sinusoidal signal at 500, 1,000, 2,000, 4,000, and 8,000 Hz, with loudness levels of 2, 7, 14, and 28 sones. We used the RMSE to quantify the difference between the sharpness values obtained from the four sharpness models and the human data of sharpness \cite{Zwicker}. The sharpness values were calculated by taking the time-averaged sharpness from the proposed sharpness models.

\begin{figure}[!tb]
    \begin{center}
        \includegraphics[width=0.5\textwidth]{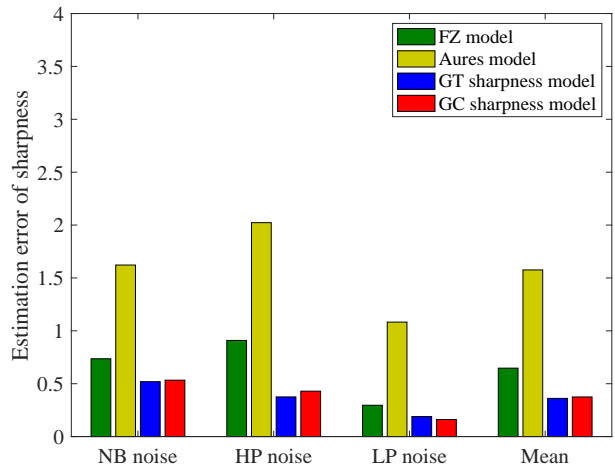}
        \caption{Estimation error of four sharpness models (FZ, Aures, proposed GT sharpness, and proposed GC sharpness models) for different types of noise}
        \label{fig:SH_R_N}
    \end{center}
\end{figure}

Figure \ref{fig:SH_R_N} shows the results of sharpness for different types of noise, where the vertical axis represents the RMSE between the results of the human data of sharpness and the values obtained from the four sharpness models. The RMSEs of the proposed sharpness models was found to be lower than those of the FZ and Aures models. Furthermore, the sharpness values obtained from the proposed sharpness models were found to be in close agreement with the human data of sharpness.

\begin{figure}[!tb]
    \begin{center}
        \includegraphics[width=0.5\textwidth]{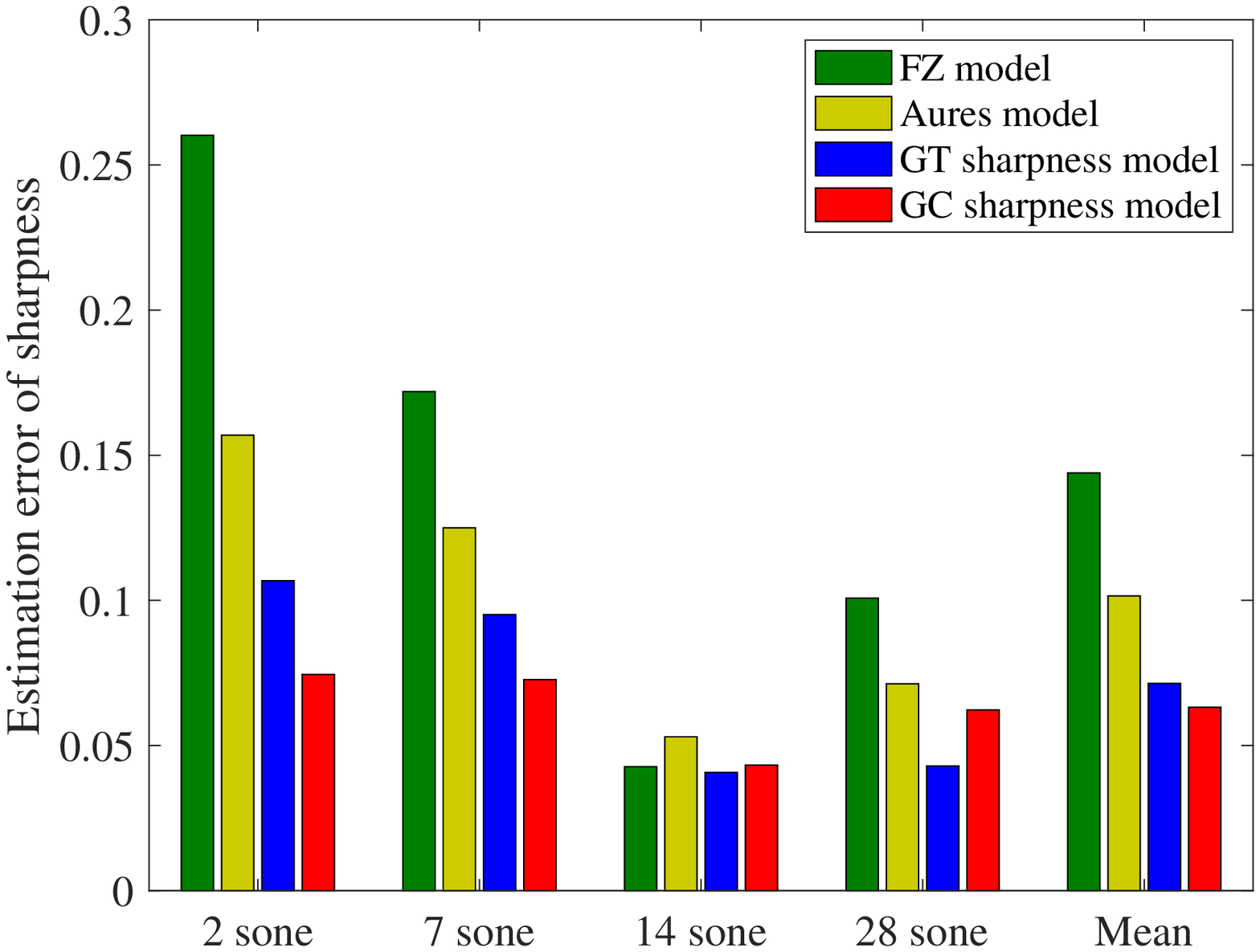}
        \caption{Estimation error of four sharpness models (FZ, Aures, proposed GT sharpnes, and proposed GC sharpness models) for different loudness}
        \label{fig:SH_R_L}
    \end{center}
\end{figure}

Figure \ref{fig:SH_R_L} shows the results of sharpness for different loudness levels, where the vertical axis represents the RMSE between the human data of sharpness and the values obtained from the four sharpness models. The RMSEs of the proposed sharpness models was found to be lower than those of the FZ and Aures models. Additionally, the sharpness values obtained from the proposed sharpness models were found to be in close agreement with the human data of sharpness.

\subsection{Evaluation of roughness}
Both versions of the proposed roughness model (GT and GC roughness models) were compared with Widmann \& Fastl’s computational model of roughness (WF model) \cite{Widmann} with Daniel \& Weber’s computational model of roughness (DW model) \cite{DanielR} in terms of modulation perception. The perceived level of roughness varies with modulation frequency, sound-pressure level, center frequency, and modulation depth. Signals that incorporate variations in these parameters were used in this evaluation.

We first assessed roughness when the modulation frequency was varied using amplitude-modulated (AM) and frequency-modulated (FM) signals. The AM signal was a 1,000-Hz sinusoidal signal with 0.2-sec duration with 100\% amplitude modulation at modulation frequencies of 10, 20, 30, 40, 50, 60, 70, 80, 90, 100, and 200 Hz. The FM signals were frequency modulated at 10, 20, 30, 40, 50, 60, 70, 80, 90, 100, and 200 Hz with a frequency deviation of 700 Hz using a sinusoidal signal of 0.2 sec at 1,500 Hz. The sound-pressure level of these signals was set at 70 dB.

We then assessed roughness when the sound-pressure level was varied using AM and FM signals. The AM signal was a 1,000-Hz sinusoidal signal with 0.2-sec duration with 100\% amplitude modulation at modulation frequencies of 70 Hz. The FM signals were frequency modulated at 70 Hz with a frequency deviation of 700 Hz using a sinusoidal signal of 0.2 sec at 1,500 Hz. The sound-pressure level of these signals was set at 40, 50, 60, 70, and 80 dB.

Next, we assessed the roughness of AM signals with varying modulation frequencies and carrier frequencies. The stimuli were sinusoidal signals of 1,000, 2,000, 4,000, and 8,000 Hz, with 0.2-sec duration with 100\% amplitude modulation at modulation frequencies of 10, 20, 30, 40, 50, 60, 70, 80, 90, 100, and 200 Hz. The sound-pressure level was set at 60 dB.

Finally, we assessed the roughness of AM signals with varying degrees of modulation. The stimuli consisted of sinusoidal signals of 1,000 Hz with 0.2-sec duration with amplitude modulation at 0, 10, 20, 30, 40, 50, 60, 70, 80, 90 and 100\% modulation and modulation frequency of 70 Hz. The sound-pressure level was set at 60 dB.

The roughness values were calculated by taking the time averaged roughness from the proposed roughness models. The RMSE was used to evaluate the discrepancy between the human data of roughness \cite{Zwicker} and predictions of the four roughness models.
\begin{figure*}[!tb]
  \center
  \includegraphics[width=1\textwidth]{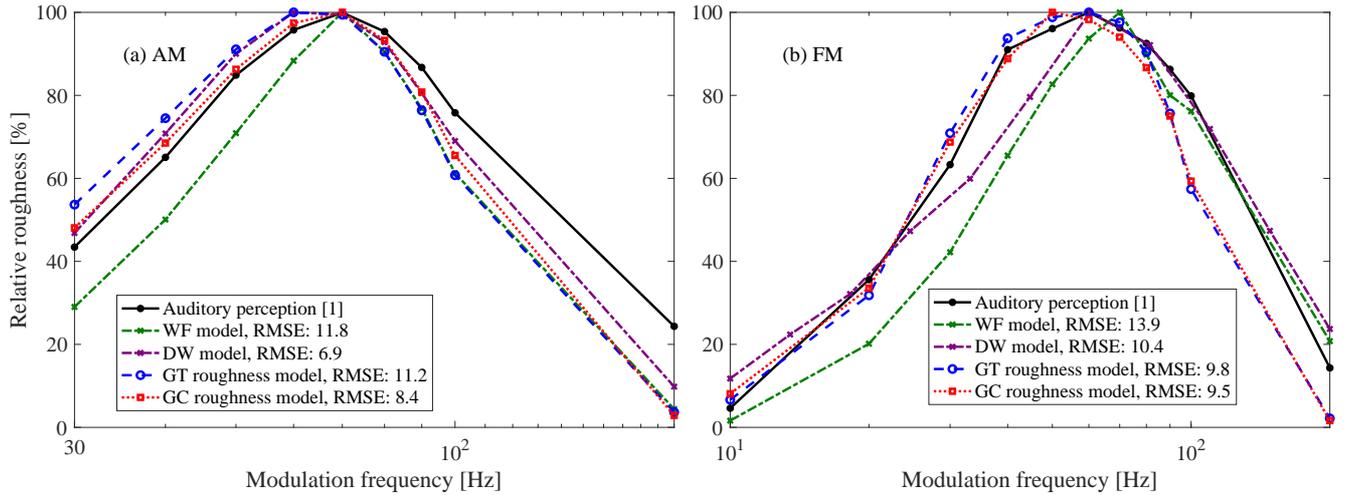}
  \caption{Relative roughness calculated using four roughness models (WF, DW, proposed GT roughness, and proposed GC roughness models) with respect to modulation frequency: (a) AM signal and (b) FM signal}
  \label{fig:RMF}
\end{figure*}
\begin{figure*}[!tb]
  \center
  \includegraphics[width=1\textwidth]{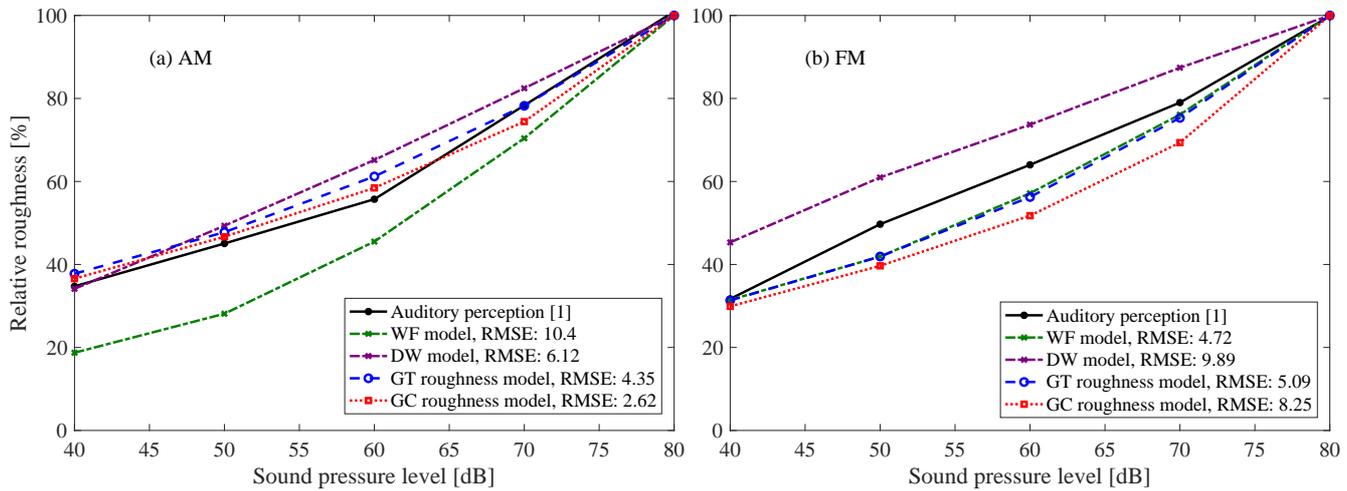}
  \caption{Relative roughness calculated using four roughness models (WF, DW, proposed GT roughness, and proposed GC roughness models) with respect to sound-pressure level: (a) AM signal and (b) FM signal}
  \label{fig:RSPL}
\end{figure*}
\begin{figure*}[!tb]
  \center
  \includegraphics[width=1\textwidth]{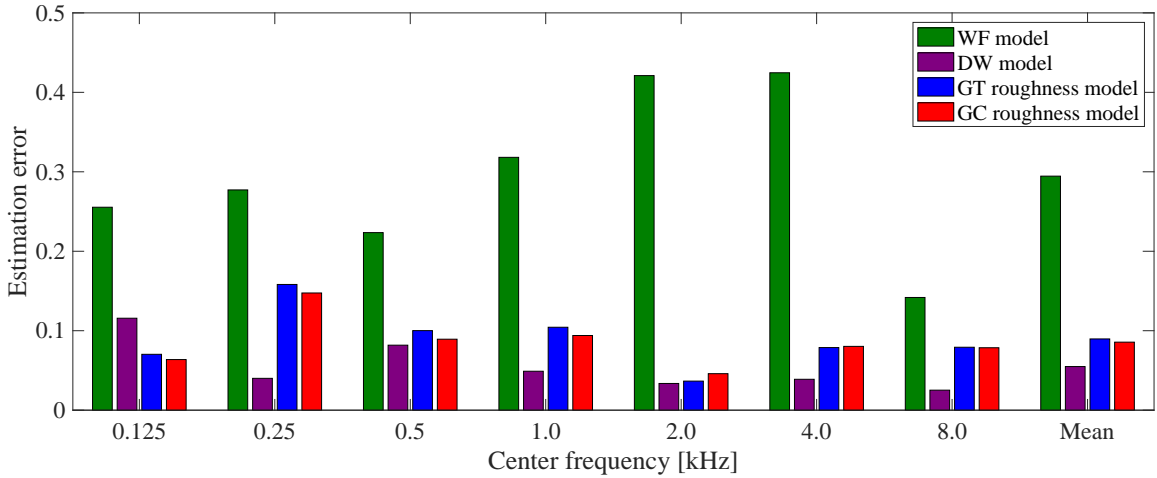}
  \caption{Relative roughness calculated by four roughness models (WF, DW, proposed GT roughness, and proposed GC roughness models) with respect to center frequency}
  \label{fig:RCF}
\end{figure*}
\begin{figure}[!tb]
  \center
  \includegraphics[width=0.5\textwidth]{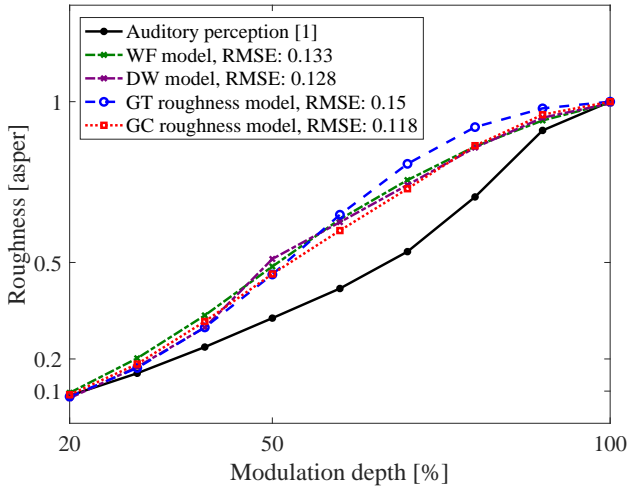}
  \caption{Roughness calculated using four roughness models (WF, DW, GT roughness, and GC roughness models) with respect to modulation depth of AM signal}
  \label{fig:Rdepth}
\end{figure}
\begin{figure*}[!tb]
  \center
  \includegraphics[width=0.95\textwidth]{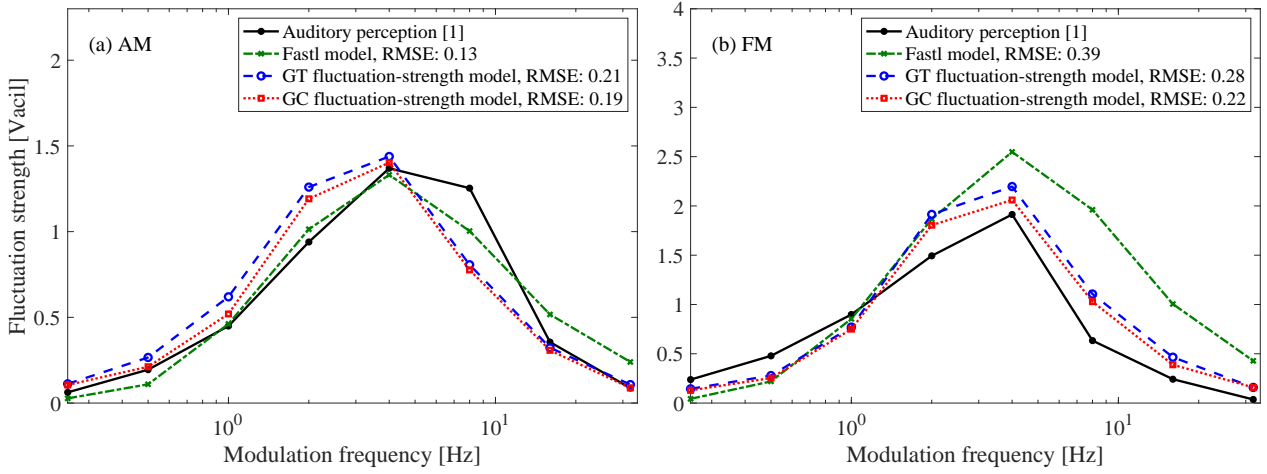}
  \caption{Fluctuation strength calculated using three fluctuation-strength models (Fastl, GT fluctuation-strength, and GC fluctuation-strength models) with respect to modulation frequency: (a) AM signal and (b) FM signal}
  \label{fig:FMF}
\end{figure*}
\begin{figure*}[!tb]
  \center
  \includegraphics[width=0.95\textwidth]{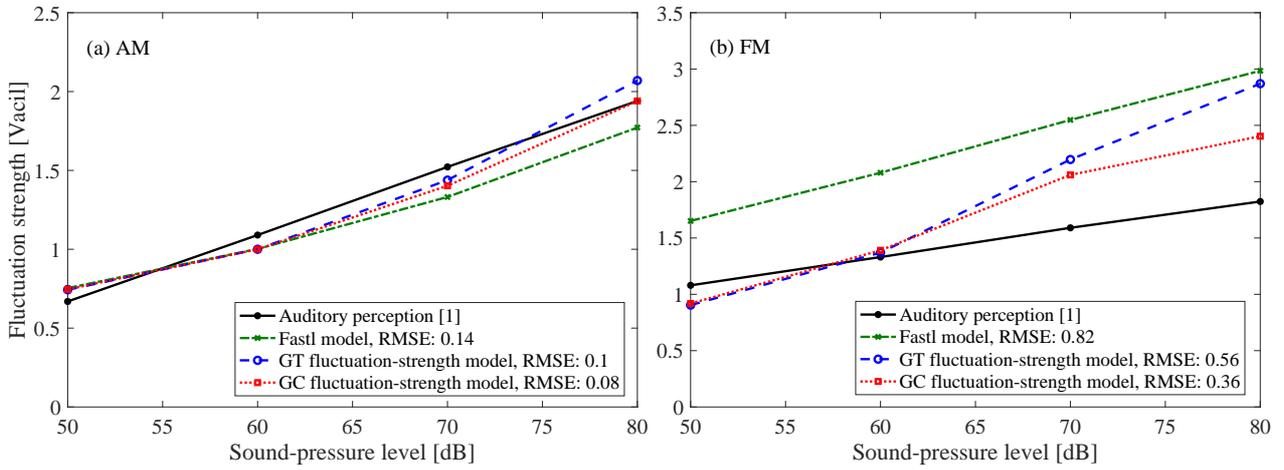}
  \caption{Fluctuation strength calculated using three fluctuation-strength models (Fastl, GT fluctuation-strength, and GC fluctuation-strength models) with respect to sound-pressure level: (a) AM signal and (b) FM signal}
  \label{fig:FSPL}
\end{figure*}
\begin{figure}[!tb]
  \center
  \includegraphics[width=0.5\textwidth]{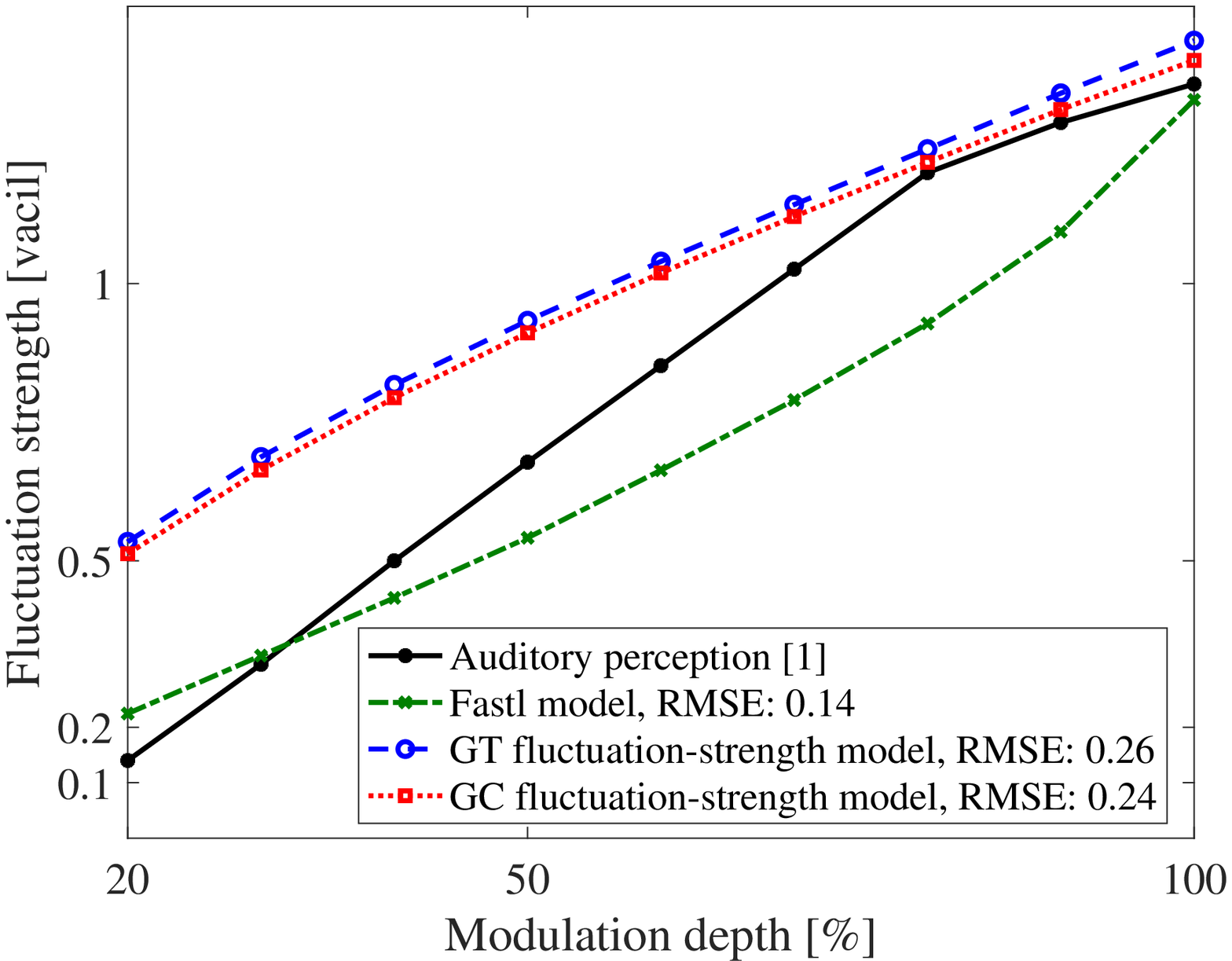}
  \caption{Fluctuation strength calculated using three fluctuation strength models (Fastl, GT fluctuation-strength, and GC fluctuation-strength models) with respect to modulation depth of AM signal}
  \label{fig:Fdepth}
\end{figure}

Figure \ref{fig:RMF} shows the roughness results obtained from the four models at various modulation frequencies for both AM and FM signals. The modulation frequency is represented on the horizontal axis and the normalized roughness on the vertical axis. The solid black line signifies the human data of roughness \cite{Zwicker}.

Figure \ref{fig:RMF}(a) shows that the WF model estimated the roughness at modulation frequencies below 70 Hz lower than the human data of roughness, while the other models estimated values that are close to the human data of roughness. The GT roughness model estimated lower results for modulation frequencies above 70 Hz compared with the human data of roughness. The computational model with the smallest estimation error was the DW model, followed by the GC roughness, GT roughness, and WF models.

Figure \ref{fig:RMF}(b) shows that the estimation results of the GC roughness model had a peak at 50 Hz. The model with the smallest estimation error was the GC roughness model, followed by the GT roughness, DW, and WF models.

Figure \ref{fig:RSPL} shows the results of roughness estimated from the four models at different sound-pressure levels for AM and FM signals. The horizontal axis is the sound-pressure level, and the vertical axis is the normalized roughness. The solid black line signifies the human data of roughness.

Figure \ref{fig:RSPL}(a) shows that the results of the WF model were estimated lower than the human data of roughness. The model with the smallest estimation error was the GC roughness model, followed by the GT roughness, DW, and WF models.

Figure \ref{fig:RSPL}(b) shows that the estimation results of the WF, GC roughness, and GT roughness models were lower than those of the listening experiment on roughness, while the estimation results of the DW model were higher than those of the human data of roughness. The model with the smallest estimation error was the WF model, followed by the GT roughness, GC roughness, and DW models.

Figure \ref{fig:RCF} shows the estimation errors of the four models for center frequency versus the human data of roughness. The horizontal axis is the center frequency, and the vertical axis is the estimation error. The model with the smallest average estimation error for all center frequencies was the GC roughness model, followed by the DW, WF, and GT roughness models.

Figure \ref{fig:Rdepth} shows the estimation results of roughness as a function of modulation depth. The horizontal axis is modulation, and the vertical axis is roughness. The solid black line signifies the human data of roughness. The four models estimated roughness higher than the human data of roughness. The model with the smallest estimation error was the GC roughness model, followed by the DW, WF, and GT roughness models.

\subsection{Evaluation of fluctuation strength}
Both versions of the proposed fluctuation-strength model (GT and GC fluctuation-strength models) were compared with the previous fluctuation-strength model (Fastl model) \cite{FastlF}. The perceived level of fluctuation strength varies with modulation frequency, sound-pressure level, and modulation depth. Hence, signals that incorporate variations in these parameters were used in the evaluation.

We first assessed the fluctuation strength when the modulation frequency was varied using AM and FM signals. The AM signal was a 1,000 Hz sinusoidal signal with 4-sec duration and 100\% amplitude modulation at modulation frequencies of 0.25, 0.5, 1, 2, 4, 8, 16, and 32 Hz. The FM signals were frequency modulated at 0.25, 0.5, 1, 2, 4, 8, 16, and 32 Hz with a frequency deviation of 700 Hz using a sinusoidal signal of 4 sec at 1,500 Hz. The sound-pressure level of these signals was set at 70 dB.

We then assessed the fluctuation strength of AM signals with varying sound-pressure levels. The stimuli were a 1,000 Hz sinusoidal signal with 4-sec duration with 100\% amplitude modulation at modulation frequencies of 4 Hz. The sound-pressure level was set at 50, 60, 70, and 80 dB.

Next, we assessed the fluctuation strength of AM signals with varying modulation depths. The stimuli consisted of sinusoidal signals of 1,000 Hz with 0.2-sec duration and amplitude modulation at 0, 10, 20, 30, 40, 50, 60, 70, 80, 90 with 100\% modulation and a modulation frequency of 4 Hz. The sound-pressure level was set at 70 dB.

The fluctuation-strength values were calculated by taking the time-averaged fluctuation strength from the proposed fluctuation-strength models. The RMSE was used to evaluate the discrepancy between the human data of fluctuation strength \cite{Zwicker} and predictions of the three models.

Figure \ref{fig:FMF} shows the results of the estimated fluctuation strength from the three models at different modulation frequencies for AM and FM signals. The horizontal axis is the modulation frequency, and the vertical axis is the fluctuation strength. The solid black line signifies the human data of fluctuation strength.

Figure \ref{fig:FMF}(a) shows that the proposed fluctuation-strength models estimated the fluctuation strength at modulation frequencies below 4 Hz more highly than the human data of fluctuation strength. However, they estimated the strength at modulation frequencies higher than 4 Hz. The model with the smallest estimation error was the Fastl model, followed by the GC fluctuation-strength, and GT fluctuation-strength models.

Figure \ref{fig:FMF}(b) shows that the estimation results of the Fastl model were higher than those of the human data of fluctuation strength for modulation frequencies above 4 Hz. The model with the smallest estimation error was the GC fluctuation-strength model, followed by the GT fluctuation-strength and Fastl models.

Figure \ref{fig:FSPL} shows the results of the estimated fluctuation strength from the three models at different sound-pressure levels for AM and FM signals. The horizontal axis is the sound-pressure level, and the vertical axis is the fluctuation strength. The solid black line signifies the human data of fluctuation strength.

Figure \ref{fig:FSPL}(a) shows that the estimated values of all the models were approximately the same as the human data of fluctuation strength. The model with the smallest estimation error was the GC fluctuation-strength model, followed by the GT fluctuation-strength and Fastl models.

Figure \ref{fig:FSPL}(b) shows that the estimation results of the WF model were higher than those of the human data of fluctuation strength. The model with the smallest estimation error was the GC fluctuation-strength model followed by the GT fluctuation-strength and Fastl models.

Figure \ref{fig:Fdepth} shows the estimation results of fluctuation strength versus modulation depth. The horizontal axis is modulation, and the vertical axis is fluctuation strength. The solid black line signifies the human data of fluctuation strength. The model with the smallest estimation error was the Fastl model followed by the GC and GT fluctuation-strength models.

\section{Conclusion}
We found that the proposed loudness method can be regarded as a time-domain method for calculating loudness based on ISO 532-2 because the RMSEs are very small. The evaluation of the proposed method showed that the calculated loudness was comparable to the loudness specified in ISO 532-2; therefore, the use of the GTFB or GCFB did not significantly affect the loudness calculation. In particular, the proposed GC loudness method was found to calculate loudness closer to those of the Moore-Glasberg method than the proposed GT loudness method. This is because the filter shape of the GCFB is similar to that of the roex auditory filter.

The output of the three proposed computational SQM models based on ISO 532-2 using the proposed loudness method was in agreement with the human data of the three SQMs \cite{Zwicker}. Comparing the estimation errors of the conventional and proposed computational SQM models, the proposed models had lower estimation errors, suggesting that the difference between the Bark and ERB measures contributed to the estimation of the SQMs. The proposed computational SQM models using the proposed GC loudness method helped explain the change in SQMs in response to changes in sound-pressure level.

Future work should involve developing a model that can process time-varying sounds using the compressive gammachirp auditory filter \cite{roexgcfb,CGammachirp} or dynamic-compressive gammachirp auditory filter \cite{DCGammachirp} as auditory filters.

\vspace{6pt}
\noindent
\textbf{Funding:} This work was supported by the SCOPE Program of the Ministry of Internal Affairs and Communications (Grant No. 201605002) and JSPS-NSFC Bilateral Programs (Grant number: JSJSBP120197416). This research was also supported by a Fund for the Promotion of Joint International Research (Fostering Joint International Research (B)) (20KK0233), from MEXT.



\begin{thebibliography}{00}

\bibitem[Zwicker(2010)]{Zwicker}
Fastl H and Zwicker E. Psycho-acoustics facts and models. Springer, La Vergne, TN USA, 2010.

\bibitem[Aures(1985)]{Aures}
Aures W. Berechnungsverfahren für den sensorischen wohlklang beliebiger schallsignale.
{\em Acta Acustica united with Acustica} 1985:130--141. German.

\bibitem[Nykanen(2009)]{Nykanen}
Nyk\"{a}nen A and Sirkka A. Specification of component sound quality applied to automobile power windows.
{\em Applied Acoustics} 2009;70:813--820. https://doi.org/10.1016/j.apacoust.2008.09.015

\bibitem[Kwon(2018)]{Kwon}
Kwon G, Jo H, and Kang J Y. Model of psychoacoustic sportiness for vehicle interior sound: Excluding loudness. {\em Applied Acoustics} 2018;136:16--25. https://doi.org/10.1016/j.apacoust.2018.01.027

\bibitem[Lionello(2020)]{Lionello} 
Lionello M, Aletta F, and Kang J. A systematic review of prediction models for the experience of urban soundscapes.
{\em Applied Acoustics} 2020;170. https://doi.org/10.1016/j.apacoust.2020.107479

\bibitem[Mian(2022)]{Mian} 
Mian T, Choudhary A, and Fatima S. An efficient diagnosis approach for bearing faults using sound quality metrics.
{\em Applied Acoustics} 2022;195. https://doi.org/10.1016/j.apacoust.2022.108839

\bibitem[5321(2017)]{5321}
ISO 532-1: 2017, Acoustics ― Methods for calculating loudness ― Part 1: Zwicker method.

\bibitem[5322(2017)]{5322}
ISO 532-2: 2017, Acoustics — Methods for calculating loudness - Part 2: Moore-Glasberg method.


\bibitem[Moore(2013)]{Moore}
Moore C J B. An introduction to the psychology of hearing. Brill, Leiden, Boston, 2013.

\bibitem[Swift(2017)]{sharpnessERB}
Swift S H and Gee L K. Extending sharpness calculation for an alternative loudness metric input. {\em Journal of the Acoustical Society of America} 2017;142. https://doi.org/10.1121/1.5016193

\bibitem[Gammatone(1988)]{Gammatone}
Patterson R D, Nimmo-Smith I, Holdsworth J, and Rice P. An efficient auditory filterbank based on the gammatone function.
{\em A meeting of the IOC Speech Group on Auditory Modelling at RSRE} 1988.

\bibitem[Unoki(1999)]{Gammachirp}
Irino T and Unoki M. An analysis/synthesis auditory filterbank based on an IIR implementation of the gammachirp.
 {\em Journal predictionstical Society of Japan} 1999;20(6):397--406. https://doi.org/10.1250/ast.20.397

\bibitem[Isoyama(2020)]{GTLoudnessFA}
Isoyama T, Kidani S, and Unoki M. Modeling of sound quality metrics using gammatone and gammachirp filterbanks.
{\em Proc. Forum Acusticum 2020} 2020;2731--2735. https://doi.org/10.48465/fa.2020.0701

\bibitem[Isoyama(2021)]{GTLoudness}
Isoyama T, Kidani S, and Unoki M. Computational models of sharpness and fluctuation strength using loudness models composed of gammatone and gammachirp auditory filterbanks.
{\em Journal of Signal Processing} 2021;25(4):141--144. https://doi.org/10.2299/jsp.25.141

\bibitem[Bismarck(1974)]{Bismarck}
Bismarck V G. Sharpness as an attribute of the timbre of steady sounds.
{\em Acta Acustica united with Acustica} 1974;159--172.

\bibitem[Aures(1985)]{AuresR}
Aures W. Ein berechnungsverfahren der rauhigkeit.
{\em Acta Acustica united with Acustica} 1985;58:268--281. German.

\bibitem[Daniel(1997)]{DanielR}
Daniel P and Weber R. Psychoacoustical roughness: implementation of an optimized model.
{\em Acta Acustica united with Acustica} 1997;113--123.

\bibitem[Duisters(2005)]{DuistersR}
Duisters R. The modeling of auditory roughness for signals with temporally asymmetric envelopes. Technische Universiteit Eindhoven 2005.

\bibitem[Fastl(1990)]{FastlR}
Fastl H. The hearing sensation roughness and neuronal responses to AM-tones.
{\em Hearing Research} 1990;46:293--296.

\bibitem[Widmann(1998)]{Widmann}
Widmann U and Fastl H. Calculating roughness using time-varying specific loudness spectra.
{\em Proc. Sound quality symposium 98} 1998;55--60.

\bibitem[Vecchi(2016)]{Vecchi}
Vecchi A O, León R G, and Kohlrausch A. Modelling the sensation of fluctuation strength.
{\em Proc. Meetings on Acoustics} 2016;28:050005. https://doi.org/10.1121/2.0000410

\bibitem[Fastl(1982)]{FastlF}
Fastl H. Fluctuation strength and temporal masking patterns of amplitude-modulated broadband noise. {\em Hearing Research} 1982:8;59--69. https://doi.org/10.1016/0378-5955(82)90034-X

\bibitem[Slaney M(1993)]{IIRGammatone}
Slaney M. An Efficient Implementation of the Patterson-Holdsworth Auditory Filter Bank. Apple Computer Tech. Rep. \#35 1993.

\bibitem[T. Irino(1997)]{Gamma}
Irino T and Patterson R D. A time-domain, level-dependent auditory filter: the gammachirp.
\newblock {\em Journal of the Acoustical Society of America} 1997;101(1):412--9.
https://doi.org/10.1121/1.417975

\bibitem[unoki(2006)]{roexgcfb}
Unoki M, Irino T, Glasberg B, Moore C J B, and Patterson D R. Comparison of the roex and gammachirp filters as representations of theauditory filter. {\em Journal of the Acoustical Society of America}  2006; 120(3):1474--92. https://doi.org/10.1121/1.2228539

\bibitem[Irino(2008)]{CGammachirp}
Irino T and Pattersonb D R. A compressive gammachirp auditory filter for both physiological and psychophysical data.
{\em Journal of the Acoustical Society of America} 2008;109(5). https://doi.org/10.1121/1.1367253

\bibitem[Irino(2006)]{DCGammachirp}
Irino T and Pattersonb D R. A dynamic compressive gammachirp auditory filterbank.
{\em IEEE TRANSACTIONS ON AUDIO, SPEECH, AND LANGUAGE PROCESSING} 2006;14(6):2222--32. https://doi.org/10.1109/TASL.2006.874669

\bibitem[DIN(2009)]{din45692}
DIN 45692: 2009. Measurement technique for the simulation of the auditory sensation of sharpness. German.
\end{thebibliography}


\end{document}